# Engineering and Validating Cyber-Physical Energy Systems: Needs, Status Quo, and Research Trends

Thomas I. Strasser[1,2] and Filip Pröstl Andrén[1]

[1] Center for Energy, AIT Austrian Institute of Technology, Vienna, Austria
[2] Institute of Mechanics and Mechatronics, Vienna University of Technology, Austria
thomas.i.strasser@ieee.org

**Abstract.** A driving force for the realization of a sustainable energy supply is the integration of renewable energy resources. Due to their stochastic generation behaviour, energy utilities are confronted with a more complex operation of the underlying power grids. Additionally, due to technology developments, controllable loads, integration with other energy sources, changing regulatory rules, and the market liberalization, the systems operation needs adaptation. Proper operational concepts and intelligent automation provide the basis to turn the existing power system into an intelligent entity, a cyber-physical energy system. The electric energy system is therefore moving from a single system to a system of systems. While reaping the benefits with new intelligent behaviors, it is expected that system-level developments, architectural concepts, advanced automation and control as well as the validation and testing will play a significantly larger role in realizing future solutions and technologies. The implementation and deployment of these complex systems of systems are associated with increasing engineering complexity resulting also in increased engineering costs. Proper engineering and validation approaches, concepts, and tools are partly missing until now. Therefore, this paper discusses and summarizes the main needs and requirements as well as the status quo in research and development related to the engineering and validation of cyber-physical energy systems. Also research trends and necessary future activities are outlined.

## 1  Introduction

Renewables are key enablers in the plight to reduce greenhouse gas emissions and cope with anthropogenic global warming [16]. The intermittent nature and limited storage capabilities of renewables culminate in new challenges that power system operators have to deal with in order to regulate power quality and ensure security of supply [7]. At the same time, the increased availability of advanced automation and communication technologies provides new opportunities for the derivation of intelligent solutions to tackle the challenges [11, 24]. Previous work has shown various new methods of operating highly interconnected power grids, and their corresponding components, in a more effective way. As a consequence of these developments, the traditional power system is being transformed into a Cyber-Physical Energy System (CPES) [31]. The electric energy system is therefore moving from a single system to a system of systems.

While reaping the benefits that come along with those intelligent behaviours, it is expected that system-level developments, architectural concepts, advanced automation and control, innovative Information and Communication Technology (ICT) as well as the validation and testing will play a significantly larger role in realizing future solutions and technologies [11, 24]. The implementation, validation, and deployment of these complex systems of systems are associated with increasing engineering complexity resulting also in increased total life-cycle costs. Proper engineering and validation approaches, concepts, and tools are partly missing until now [26].

The aim of this paper is to discuss and summaries the main needs and requirements as well as the status quo in research and development related to the engineering and validation of CPES. Also research trends and necessary future activities are outlined.

The remaining parts of the paper are organized as follows: Section 2 provides a brief overview of the desired engineering and validation process of CPES-based applications whereas in Section 3 the main problems and needs are identified. Afterwards, the status quo in research and development is analysed in Section 4 followed by a discussion of future research trends in Section 5. Finally, the main conclusions are provided in Section 6.

## 2   Desired CPES Engineering and Validation Process

As already mentioned above, the development of new application for CPES is associated with increasing engineering complexity. One main issue with this is that traditional engineering methods for power system automation were not intended to be used for applications of this scale and complexity. By combining the outcome of several publications in the last few years, a desired engineering and validation process can be proposed containing four main phases: *(i)* specification and use case design, *(ii)* automated engineering, *(iii)* validation, and *(iv)* deployment [8, 4, 21, 3, 13]. The proposed process is based on a model-driven approach and is illustrated in Fig. 1 with the main focus on providing the user with automated support throughout the whole process.

The process starts with a specification and use case design. Based on use case description methods such as the Smart Grid Architecture Model (SGAM) and IEC 62559 (also known as the IntelliGrid approach), a structured description and visualization of use cases can be defined. However, to fully take advantage of the high amount of information in the modelled use cases these description methods must be transformed into machine-readable and formal formats [2]. Formal specifications together with the high-level use case description act as the main input and thus form the basis for the following automated steps of the development process [22].

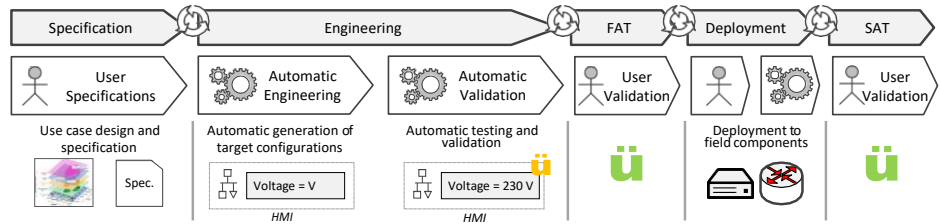

**Fig. 1.** Overview of a desired engineering and validation process of CPES [22].

The following automated steps in the process are based on the formal specifications and use case descriptions. First of all, it is possible to generate different types of configurations. This can be executable code for field devices, configurations for ICT equipment or generated setups for Supervisory Control And Data Acquisition (SCADA). Model-based or ontology based approaches from software engineering are examples of methods that can provide this kind of automation [22, 32]. Optimally it should also be possible to automatically validate the resulting generated output configurations. Automated testing for software development has already existed since several years. However, similar approaches for CPES are currently missing. When testing CPES, multiple system aspects need to be validated, such as ICT, automation/control, and security and privacy topics apart from the power system itself [3, 25]. For system tests the advantage of automated testing would be even greater than for single components.

Once validation has been successfully accomplished the last phase is deployment of the software and configurations to field devices. When the power system relies on more and more software systems for a secure and safe operation, the deployment and update process of this software gets more complicated. If updates are not applied in the correct order, or if an update is unsuccessful, this can cause the power system to operate incorrectly or even damage it or people working with it. To support the user with this stage, the development process should be able to analyze the power system state and create rollout schedules that minimizes the risk of disruptive events and errors [13, 5, 27].

## 3   Main Problems and Needs

A characteristic feature of renewable sources and customer side solutions is that they are mostly available in a decentralized way as DER [16]. From todays point of view it is nearly impossible to address all future needs and requirements in CPES and DER components. The flexible addition of new control functionalities in DER components is therefore an essential future feature [33]. For example, in Germany about 15 GW power is produced from PV systems today. The 50.2 Hz problem is a well-known fact because today it is nearly impossible to remotely update control functions and parameters of DER components to correct earlier grid code decisions [30]. Beyond purely technical solutions, changes in regulations and grid codes will also be indispensable. Consequently, the planning, management, and operation of the future power systems have to be redefined. The implementation and deployment of these complex systems of systems are associated with increasing engineering complexity resulting in increasing engineering costs.

The usage of proper methods, automation architectures, and corresponding tools offers a huge optimization potential for the overall engineering process. The starting point is detailed use case and requirements engineering. This has led to a number of recent smart grid projects where use case descriptions of corresponding applications are in the focus [9, 23]. Some promising approaches have already been developed like the IntelliGrid method [12] or SGAM [10, 28, 29] but they are mainly lacking of a formalized and computer-readable representation for engineering and validation automation. The same is true for other input specifications that are typically provided as an input to the engineering process. Also, there is no standardized way of representing the objects, e.g., controllers and power grid components, neither in the way they are depicted, nor in the semantics used in the description.

A significant amount of work has to be spent a repeated number of time *(i)* in the specification, *(ii)* the implementation, *(iii)* the validation and testing but also *(iv)* in the deployment phase. This is a very time-consuming and error-prone approach to design and test CPES applications. To conclude, the following main problems related to the engineering of CPES exist today [25, 26, 15]:

- *Rigorous engineering:* Rigorous model-based engineering concepts for CPES applications are missing or only partly available.
- *Rapidness and effort:* CPES solutions are becoming more and more complex, which results in increasing engineering efforts and costs. Therefore, it is important to improve the rapidness of traditional engineering methods.
- *Correctness:* Due to the multidisciplinary character of CPES applications this also requires the engineer to have an expert knowledge in each discipline. This is often not the case, which increases the risk of human errors.
- *Handling legacy systems:* Power grid operators expect a long service life of all components in their systems. Available proprietary automation solutions in smart grid systems prevent efficient reuse of control software.
- *Geographical dispersion:* The distribution of components over large geographical areas requires special attention. New ICT approaches and wide-area communication are needed.
- *Interoperability:* Interoperability is a critical issue in CPES applications. This must be assured on all levels, from specifications over implementation, to deployment and finally during operation. Also components from different manufacturers must be handled, which requires a manufacturer independent method.
- *Real-time constraints:* Some applications may enforce real-time constraints on hardware, software and networking. Performance management is often not adequately addressed in existing engineering processes.

- *Scalability:* Current engineering methods are focusing on the development for a single system. With the introduction of new intelligent grid components and DER these methods must be able to handle not only a single system, but a system of systems.
- *Reference scenarios:* Common and well understood reference scenarios, use cases, and test case profiles for CPES need to be provided to power and energy systems engineers and researchers; also, proper validation benchmark criteria and key performance indicators as well as interoperability measures for validating CPES solutions need to be developed, extended, and shared with engineers and researchers.
- *Deployment support:* An easy, secure, and trustable deployment of CPES applications to a large amount of corresponding components and devices is necessary.
- *Standardization:* A harmonization and standardization of multi-domain CPES-based engineering and validation approaches as well as corresponding and testing procedures is required.

If the information gathered during the use case description phase can also be used directly in an automated method, the development effort can substantially be decreased. By collecting the use case information in a formal model this can be used for direct automatic code generation. The result can be executable code for field devices, communication configurations as well as HMI configurations. Moreover, an automated approach also has the potential to de-crease implementation errors and at the same time increase the software quality [18].

The availability of both a formal specification of CPES use cases and the possibility to automatically generate target configurations also enables the use of automated testing. For CPES this can be pure software tests but can also be a combination of software, hardware and simulations. When automatic validation errors can be detected at an early stage, this will increase the overall quality and mitigate the development risk. The quality of CPES applications is also very much dependent on the experience and knowledge of the responsible engineer(s). Furthermore, the electric energy system is becoming a multi-domain system. This requires knowledge not only about the power system but also about ICT, control, and automation issues. In order to benefit from the knowledge of experienced engineers, machine learning can be used to reason about user design experience.

In the following sections the status quo in research and development is analysed and an outlook about potential future research directions are provided.

## 4  Status Quo in Research and Development

Promising approaches and research results categorized according to the above outlined process (i.e, see Fig. 1) related to the engineering and validation of CPES applications and solutions are briefly summarized and discussed below.

### 4.1 Specification Phase

An approach for specifying and modelling CPES was introduced first with IntelliGrid [12] and later refined by SGAM [17]. In principle, SGAM provides a set of concepts, viewpoints, and a mapping method and thus a structured approach for smart grid architecture development. It allows depicting various aspects of smart grid artefacts (information, communication, etc.) and supports identification and coordination of elements on different levels. It also facilitates the identification of interoperability issues [10, 29, 28]. The basis for SGAM is a three-dimensional frame consisting of domains, zones and layers. The use of the SGAM modelling approach combined with the IntelliGrid use case template provides a powerful methodology for smart grid use case and application development. The design steps as outlined in Fig. 2 define the structured approach [17]:

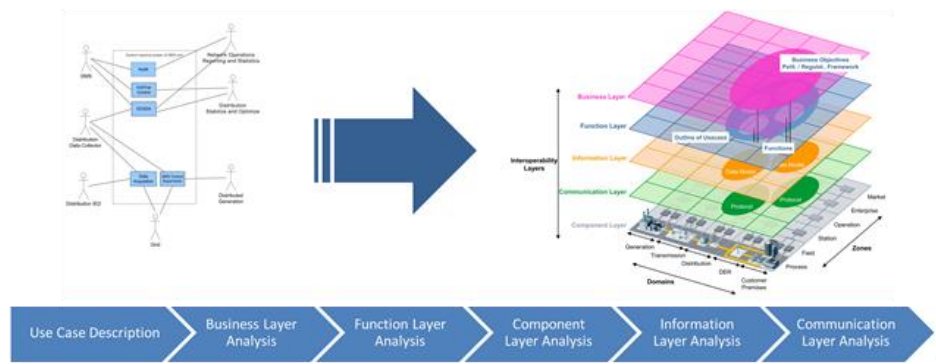

**Fig. 2.** SGAM-based design process for smart grid use cases and applications [17].

### 4.2 Engineering Phase

**Model-Based Engineering:** In order to provide tool support for the methodology illustrated in Fig. 2, Dänekas et al. developed the "SGAM Toolbox", which is an extension to the Enterprise Architect (EA) software [6]. This has the advantage that standard UML modelling tools, such as sequence or activity diagrams can be used by the engineering since these are standard parts of the EA tool. With the SGAM Toolbox, support is provided to cover modeling of all steps in the SGAM use case methodology. Additionally, due to the already included code generation capabilities of EA, it is also possible to generate certain code components based on the models made by the SGAM Toolbox. One drawback, is that the code generation is intended for general-purpose applications and thus no extended generation support for other configurations (e.g., communication or HMI configurations) is provided [21].

Another work, also based on the SGAM Toolbox, by Knirsch et al. implements a model-driven assessment of privacy issues [14]. To do this, the analysis is based on modeled data flows between actors and secondly, an assessment can be made during design time to study what impact it has on the modeled use case. A related approach was also developed by Neureiter et al. [19], where they use the SGAM models together with the "Guidelines for Smart Grid Cybersecurity" from the National Institute of Standards and Technology (NIST) [1]. By including cybersecurity issues directly into the modeling of smart grid use cases they show how this be used to study potential security implications, which can serve as a basis for further implementations.

Another holistic approach for rapid development of intelligent power utility automation applications was presented by Andrén et al. [2, 21]. In [2], the outline of a Model-Driven Engineering (MDE) approach for automated engineering of smart grid applications was presented. This was later complemented with a Domain-Specific Language (DSL) for SGAM compliant use case design [21]. Compared to the SGAM Toolbox this work also focuses on code generation as well as generation of communication configurations. The result of these two publications is a model-based rapid engineering approach for smart grid automation applications.

**Ontology-Based Engineering:** Ontologies are a way to abstract domain specific information, usually with the aim to represent objects, types, and their semantic relations in a formal machine-readable way. Additionally, a set of inference rules and restrictions on relations can be defined to support semantic interpretation. An approach based on ontologies was used by Zanabria to model multi-functional Energy Storage System (ESS) applications and to handle inconsistencies derived from the overlapping of corresponding applications [32]. Focus was put on handling Energy Management System (EMS) application and for this, the EMS Ontology (EMSOnto) was created [32]. An overview of the main EMSOnto concepts are shown in Figure 3.

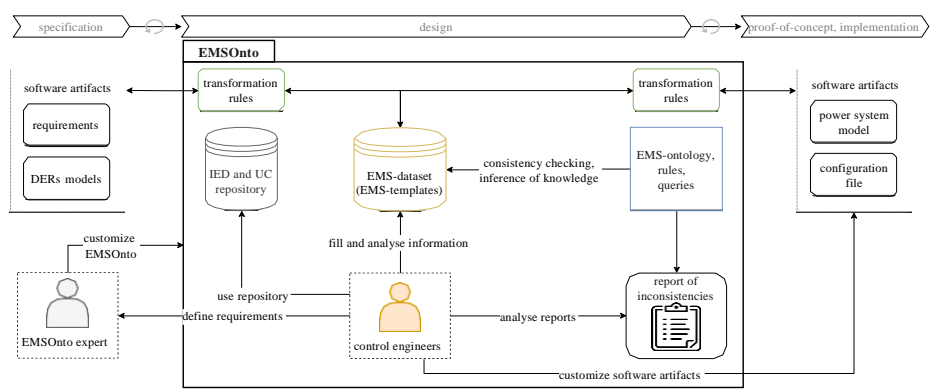

**Fig. 3.** Overview of the EMSOnto engineering process [32].

The core ontology defines the structure of a database (EMS-database) used to gather relevant information for the design of the EMS functions. This information can be available at the specification stage or it can be provided manually by control engineers. The whole database is checked and validated against terminological axioms contained in the EMS-ontology. A reasoner engine uses the validated database to execute complex axioms and logical rules in order to deduce new information. This reasoning is used on the information in the database in order to identify inconsistencies and also generate handling proposals of any identified conflicts. This information is included within reports that are presented to the user, who analyses the reports and can decide on improvements and modifications within the EMS design. The referred process is recursively executed in order to achieve an error-free design [32].

### 4.3 Validation and Testing Phase

**ERIGrid Holistic Validation Procedure:** The specification and execution of tests or experiments is central to any assessment approach. The holistic testing procedure proposed in ERIGrid [25] aims to unify the approach to testing across different research infrastructures, different testbed types and to facilitate multi-domain testing. A central element of the ERIGrid approach is the Holistic Test Description (HTD) method [3]; it defines a number of specification elements for any test, to be identified independently of the particular assessment methodology. The specification elements comprise three main levels of specification as outlined in Fig. 4:

- *Test Case (TC):* a concise formulation of the overall test objectives and purpose of investigation, specifying the object under investigation, separating system structure from functions (system under test vs. functions) and identifying test criteria.
- *Test Specification (TS):* a detailed description of the test system, control parameters and test signals to address a specific subset of the test criteria.
- *Experiment Specification (ES):* how is the test specification realized in a given testbed? What compromise had to be made in the representation with respect to the test specification?

The method is supported by structured templates to facilitate the harmonized recording of test descriptions as outlined also in Fig. 4. In addition it also guides the test engineer in the identification of validation systems configurations (i.e., from a generic to a specific and finally to a lab-based representation).

**JRC Interoperability Testing Methodology:** The interoperability in CPES between different components and devices is an important issue. Testing the interoperability in this domain requires producing detailed test cases describing how the different players, services, and components are intended to interact with each other. A systematic approach for developing CPES-based interoperability tests has the potential to facilitate the development of new solutions and applications. Therefore, the Joint Research Center (JRC) of the European Commission (EC) has developed a corresponding methodology.

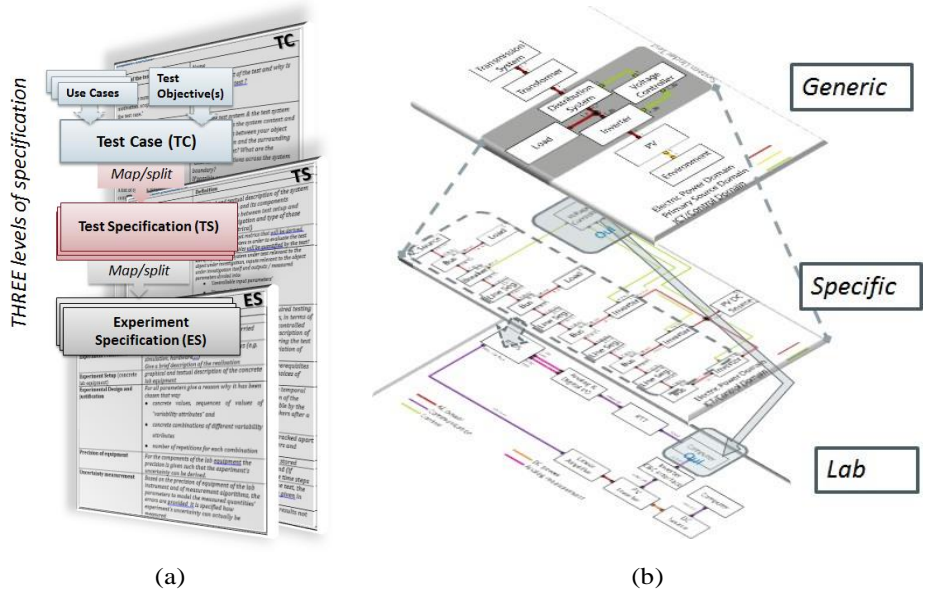

(a)                                                 (b)

**Fig. 4.** Overview of the ERIGrid holistic test description: (a) holistic testing methodology and corresponding templates, (b) validation system configurations on different levels (adapted from [3]).

It helps the CPES engineer in a structured way to create interoperability testing Use Cases, Basic Application Profiles (BAP) and Basic Application Interoperability Profiles (BAIOP) and it is used mainly as a common framework for interoperability testing. Five main steps have been identified which are *(i)* Use Case creation, *(ii)* Basic Application Profile (BAP) creation, *(iii)* Basic Application Interoperability profile (BAIOP) creation, *(iv)* Design of Experiments (DoE) as well as the *(v)* Testing and Analysis of Experiments (AE). Fig. 5 provides a brief overview of the overall approach.

Each stage allows the engineer to select certain features then used in the subsequent stage. During the completion of all stages, the developer can select relevant standards, their options, test beds with all qualified and test equipment as well their attributes or functions used during the testing.

### 4.4 Deployment Phase

Especially for future CPES, where a large number of functions and services interact with each other within a complex dynamic system, processes for deployment and software update become more important and more complex. It is important that deployment processes can ensure that all dependencies on all layers (power system, communication, information) are fulfilled. Furthermore, these processes also must be resilient against faults and attacks (internal and external). The LarGo! project aims at enabling mass roll-out of smart grid applications for grid and energy management by defining a seamless, safe and secure application deployment process for the grid and customer domain [13].

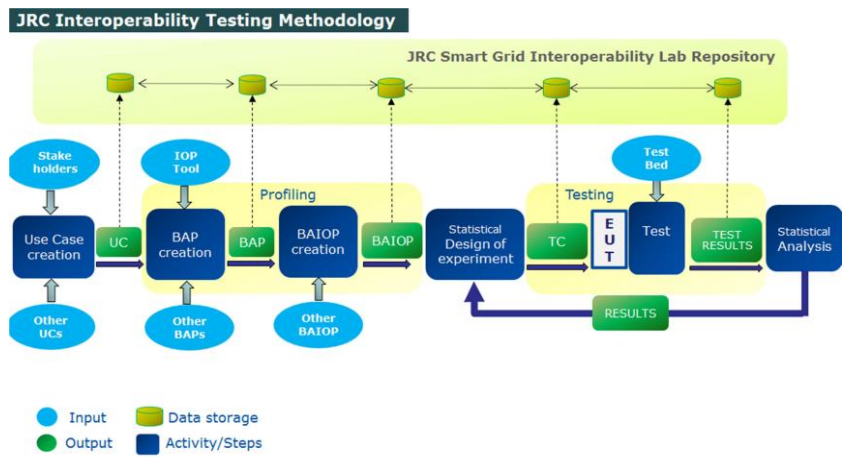

**Fig. 5.** Overview of the JRC interoperability testing approach [20].

In the LarGo! project two main approaches are followed. First, a new and inter-domain software deployment process is designed to monitor multiple conditions (e.g., timing, power system state, business and customer issues, weather conditions) to ensure that the deployment of software results in the desired overall system behavior. To enable the operator to create an optimal rollout schedule, a tool helps the operator with a guided and assisted process. This is combined with a knowledge-based systems to provide and resolve necessary conditions and requirements [13]. The second part of LarGo! complements the deployment process with secure and resilient system design. This includes developments of resilient control applications that can handle disruptive events, such as faults and attacks, and also cyberssecurity concepts with the goal to detect root-causes of errors after a software update (i.e., was the error caused by an attack during the deployment process) [5, 27].

## 5   Discussion of Future Research Needs

In fact, current research show that already many aspects needed for a better engineering and validation of CPES are available. Nevertheless, many issues are still open and since the advancement of CPES technologies is still ongoing, new needs are constantly appearing. In Section 3 several problems and needs were pointed out. Some of these can already—at least partly—be solved by using the approaches presented in this paper. The following list is an attempt to summarize the above presented approaches, show how the identified problems and needs can be solved, and to point out possible research directions that still needs to be explored:

- *Harmonization of existing approaches*: As already shown in this paper, there are already quite many approaches that attempt to improve the engineering and validation of CPES. However, it is also clear that these approaches all have their own focus and cannot cover the whole development process, as it is depicted in Fig. 1. Therefore, one important midterm goal for future research will be to harmonize already existing approaches. This will directly provide better support for *rigorous engineering*, *rapidness and effort*, *correctness*, and *deployment support* of CPES applications since many approaches already focus on these. Furthermore, through a harmonization a better *interoperability* can be reached. A logical continuation, and a more longterm goal, will be to integrate the harmonized approaches within international *standardization* organizations.
- *Large-scale examples and scenarios*: Many of today's approaches from research are still to be proven on large-scale real world examples. More research is needed were the different methods are applied to larger use cases. This is important to solve any remaining issues regarding *scalability*. Associtated is also the development of *reference scenarios* that can help to test and benchmark future engineering and validation approaches as they emerge.
- *Integration with traditional engineering approaches*: Components and systems for the power system domain are known for their long lifespan. Due to this, it will also be necessary for any new and future development process to integrate with already existing traditional engineering approaches. This is one the one hand necessary in order to increase the acceptance of the new development approaches On the other hand it will also be needed to *handle legacy systems* and to ensure *interoperability* with older components.
- *Introducing new abstractions and modeling options*: Following the continuous development of CPES, the associated engineering and validation methods need to evolve as well. In order to handle increasingly complex systems development, new abstractions and modeling options will be needed to support engineers. *Real-time constrains*, timing [15], *geographical dispersion*, and *scalability* are examples of abstractions that are currently not supported by any of the discussed approaches. The more complex the CPES gets, the more computer-aided support—for example through machine-learning—will be needed in order to handle these issues.

Summarizing, engineering and validation support will be critical for successful development of future CPES applications. Without the proper tool support many of the tasks will require immense manual efforts and will require engineers educated in multiple domains (e.g., energy system physics, ICT, automation and control, cybersecurity). The tools available today and currently in development all provide small steps in the right direction, but more research efforts are still needed in the years to come.

## 6    Conclusions

The current electric energy system is moving towards CPES. Integration of more and more renewable energy sources require a change of tactics for the planning and operation today's power systems. New technologies, such as ICT, automation, and control systems are needed together with changing regulatory rules and a market liberalization. The implementation and deployment of these complex systems of systems are associated with increasing engineering complexity resulting also in increased engineering costs. Proper engineering and validation approaches, concepts, and tools are partly missing until now.

In this paper the most major needs for future CPES engineering and validation approaches have been discussed. Furthermore, the current state of research in this are today have been presented. The development of CPES applications has been divided into four main phases: specification, engineering, validation, and deployment. For each phase corresponding approaches have been presented, highlighting advantages and disadvantages. This is followed by a discussion on future research needs.

As a conclusion, engineering and validation support will be critical for successful development of future CPES applications. Some needs are already covered by today's approaches, but even more can be covered by harmonization of those that already exist. Current approaches will also need to introduce new abstractions and modeling possibilities once future CPES applications increase in complexity. Examples are real-time constraints and timing, which are not yet properly covered. In summary, more support can still be provided.

**Acknowledgments.** This work is partly funded by the Austrian Ministry for Transport, Innovation and Technology (bmvit) and the Austrian Research Promotion Agency (FFG) under the ICT of the Future Program in the MESSE project (FFG No. 861265), by the European Community's Horizon 2020 Program (H2020/2014-2020) in the ERIGrid project (Grant Agreement No. 654113), and by the framework of the joint programming initiative ERA-Net Smart Energy Systems' focus initiative Smart Grids Plus, with support from the European Union's Horizon 2020 research and innovation programme under grant agreement No 646039 in the LarGo! project (national funding by the Austrian Climate and Energy Fund FFG no. 857570).